\documentclass[aps,prd,showpacs]{revtex4}
\usepackage{graphicx}
\usepackage{psfig}
\usepackage{revsymb}

\begin{document}
\title{On The Transport Theory of Classical Plasma in Rindler Space}
\author{\footnotesize Soma Mitra$^{a,1}$ and Somenath Chakrabarty$^{a,2}$}
\affiliation{$^a$Department of Physics, Visva-Bharati, Santiniketan 
731235, India\\
$^1$somaphysics@gmail.com\\
$^2$somenath.chakrabarty@visva-bharati.ac.in
}
\begin{abstract}
We have obtained the Vlasov equation and Boltzmann kinetic
equation using the classical Hamilton equation and 
Poisson bracket with Rindler Hamiltonian. 
We treat the whole Universe as a statistical system with
galaxies as the point particle constituents 
in large scale structure. Since the collisions of 
galaxies are very rare phenomena, we assume that the gas with
the constituents as point galaxies satisfy Vlasov equation.
Considering the astrophysical catastrophic event, e.g., 
the generation of gravity waves
by the collisions of black holes in one of the galaxies, 
and further assuming that when
such a wave passes through the gas causes a kind of asymmetry in
mass distribution. We may call it a kind of polarization (of course
there is no mass and anti-mass). This polarization of mass distribution will
further gives rise to gravitational permittivity or dielectric
constant. We have shown that the low frequency gravity waves will
be absorbed, whereas the high frequency part will pass through the
gas of point galaxies. It is further noticed that the region in
space with extremely high gravitational field is transparent to
gravity waves. In the other part of this work, using the
Boltzmann equation and replacing the collision term by the
relaxation time approximation and further assuming a
small deviation from the equilibrium configuration of the stellar /
galactic plasma in
Rindler space, we have obtained the kinetic coefficients. For the
first time we have derived an expression for the coefficient of
gravitational flow. It has further been shown that in presence
of strong gravitational field all the kinetic coefficients become
vanishingly small, so the matter behaves like an ideal fluid, which may
happen near the event horizon of black holes.
\end{abstract}
\maketitle
\section{Introduction}
Exactly like the Lorentz transformations of space time coordinates in
the inertial system of frames, the Rindler coordinate
transformations are for the uniformly accelerated frame of
references \cite{R1,R2}. The space-time geometrical structure is called the
Rindler space.
From the references \cite{R3,R31,R32,R4,R5,R6,R7,R71}, it can very easily be
shown that the Rindler 
coordinate transformations are given by: 
\begin{eqnarray}
ct&=&\left (\frac{c^2}{\alpha}+x^\prime\right )\sinh\left (\frac{\alpha t^\prime}
{c}\right ) ~~{\rm{and}}~~ \nonumber \\
x&=&\left (\frac{c^2}{\alpha}+x^\prime\right )\cosh\left (\frac{\alpha t^\prime}
{c}\right ) 
\end{eqnarray}
Hence it is a matter of simple algebra to prove that the  inverse
transformations are given by:
\begin{equation}
ct^\prime=\frac{c^2}{2\alpha}\ln\left (\frac{x+ct}{x-ct}\right )
~~{\rm{and}}~~ x^\prime=(x^2-(ct)^2)^{1/2}-\frac{c^2}{\alpha}
\end{equation}
Here $\alpha$ indicates the uniform acceleration of the frame along
positive $x$-direction. Hence it can
very easily be shown from eqns.(1) and (2) that the square of the
four-line element changes from
\begin{eqnarray}
ds^2&=&d(ct)^2-dx^2-dy^2-dz^2 ~~{\rm{to}}~~\nonumber \\ ds^2&=&\left
(1+\frac{\alpha x^\prime}{c^2}\right)^2d(ct^\prime)^2-{dx^\prime}^2
-{dy^\prime}^2-{dz^\prime}^2
\end{eqnarray}
where the former line element is in the Minkowski  space, whereas
the later one is in Rindler space.
Hence the metric in the Rindler space can be written as
\begin{equation}
g^{\mu\nu}={\rm{diag}}\left (\left (1+\frac{\alpha x}{c^2}\right
),-1,-1,-1\right )
\end{equation}
whereas in the Minkowski space-time we have the usual form
\begin{equation}
g^{\mu\nu}={\rm{diag}}(+1, -1, -1, -1)
\end{equation}
It is therefore quite obvious that the Rindler space is also flat. The only difference from the Minkowski space is
that the frame of the observer is moving with uniform acceleration.
In $1+1$-dimension, the Rindler metric is given by
\[
g^{\mu\nu}={\rm{diag}}\left (\left (1+\frac{\alpha x}{c^2}\right
),-1\right )
\]
It has been noticed from the literature survey 
that the principle of equivalence plays an important role in obtaining the
Rindler coordinates in the uniformly accelerated frame of reference. 
According to this principle an accelerated frame in absence
of gravity is equivalent to a frame at rest in presence of 
gravity. Therefore in the present scenario, $\alpha$ may also be treated 
to be the strength of 
constant gravitational field for a frame at rest.

Now from the relativistic dynamics of special theory
of relativity, the action integral is given by \cite{R1}
\begin{equation}
S=-\alpha_0 c \int_a^b ds\equiv \int_a^b Ldt
\end{equation}
where $\alpha_0=-m_0 c$  and $m_0$ is  the
rest mass of the particle and $c$ is the speed of light in vacuum
\cite{R1}.
The Lagrangian of the particle may be written as
\begin{equation}
L=-m_0c^2\left [\left ( 1+\frac{\alpha x}{c^2}\right )^2 -\frac{v^2}{c^2}
\right ]^{1/2}
\end{equation}
where $v$ is the component of three velocity along positive
$x$-direction. Hence the component of  three 
momentum of the particle along positive $x$-direction is given by
\begin{equation}
p=\frac{\partial L}{\partial v}, ~~ {\rm{or}}
\end{equation}
\begin{equation}
p=\frac{m_0 v}{\left [ \left (1+\frac{\alpha x}{c^2} \right )^2
-\frac{v^2}{c^2} \right ]^{1/2}}
\end{equation}
Then from the definition, the Hamiltonian of the particle may be written as
\begin{equation}
H=pv-L ~~ {\rm{or}}
\end{equation}
\begin{equation}
H=m_0c^2 \left (1+\frac{\alpha x}{c^2}\right ) \left (1+
\frac{p^2}{m_0^2c^2}\right )^{1/2}
\end{equation}
Hence it can very easily be shown that in the non-relativistic approximation, the Hamiltonian is given by 
$$
H=\left (1+\frac{\alpha x}{c^2}\right ) \left (\frac{p^2}{2m_0}+m_0c^2 \right ) \eqno(11a)
$$

In this study our intention is to show that the gravity waves and the
electromagnetic waves behave differently in presence of strong
gravitational field. For the gravity waves, the refractive index
will be real and constant, therefore a plane gravity wave  will travel
in straight line without any deviation near the event horizon. Whereas it
is well established that there will be bending of  electromagnetic waves
near strong gravitational field. This bending can also be explained
by the variation of refractive index. One can show very easily (see
\cite{R1,SOSO}) that the refractive index increases towards the
increasing gravitational field. It can also be shown that for the
electromagnetic waves, gravity behaves like a refracting medium with
changing refractive index as the gravity changes in the free space. In
the second part of this investigation we have shown that in presence of
strong gravitational field all the kinetic coefficients become almost
zero. Therefore a stellar plasma in presence of strong gravitational
field will behave like an ideal fluid.

The manuscript is organized in the following manner: In the next
section considering the whole universe to be a statistical system, and
further assuming that the galaxies are point objects in large scale
structure, we
have developed a formalism to obtain the Vlasov equation satisfied
by this point objects. With a harmonic type perturbation by strong
gravity waves, we have
obtained the gravitational permittivity of the medium and studied
its variation with frequency and also with the strength of
gravitational field. In section 3 we have obtained the kinetic
coefficients for a stellar plasma using the
Boltzmann kinetic equation in Rindler space. To the best of
our knowledge such studies have not been done before.
\section{Vlasov Equation in Rindler Space}
We assume that the whole universe is a statistical system and in the
large scale structure the galaxies are considered to be the point
particles, the constituents of this statistical system. Since the
collision of galaxies is a very rare phenomena, we may assume that these
point structure galaxies satisfy Vlasov equation. We further assume
that the gas is in an average uniform gravitational field $\alpha$.
The
space time geometry we have considered is  the so called 
Rindler space. 

To obtain the Vlasov equation, we start with the well known Hamilton
equation, given by
\begin{equation}
\frac{df}{dt}=\frac{\partial f}{\partial t}+\{f,H\}
\end{equation}
where distribution function $f(x,p,t)$ is a function of space, time
and momentum coordinates of the particle. Now using  the definition of
Poisson bracket, we have 
\begin{equation}
\{f,H\}=\{f,H\}_{x,p}=\frac{\partial f}{\partial x}\frac{\partial
H}{\partial p}-\frac{\partial f}{\partial p}\frac{\partial
H}{\partial x}
\end{equation}
Therefore for the Rindler Hamiltonian $H(x,p)$, given by eqn.(11),
we have for the collision-less case ($df/dt=0$),
\begin{equation}
\frac{\partial f}{\partial t}-v\beta(x)\left (1+
\frac{p^2}{m_0^2c^2}\right )^{-1/2} \frac{\partial f}{\partial x}
-m_0\alpha \left ( 1+\frac{p^2}{m_0^2c^2}\right )^{1/2}
\frac{\partial f}{\partial p}=0
\end{equation}
This is the Vlasov equation in Rindler space satisfied by the
galaxies assumed to be point particles in large scale structure.
We further assume that the average mass of the galaxies is $m_0$
(see \cite{LP} for the electromagnetic case).
Let us now assume some kind of astrophysical catastrophic events, e.g., 
generation
of very high energy gravity waves by the collision of black holes in one
of the point galaxies. The physical picture is that the gravity waves
are coming out from one of these point particles and falls on the
collision-less plasma of galaxies. 
Although there are many harmonics present in the gravity waves, we start with a
monochromatic gravity waves and later investigate the interaction of
very low and very high frequency gravity waves on such
collision-less plasma. We assume that such high intensity gravity
waves falling on this collision-less plasma causes some kind of polarization
of mass distribution. This is of course not exactly like the electrostatic
case. It is the asymmetry of mass distribution of plasma particles
caused by the incident gravity waves. 
This polarization effect may be represented by the
deviation of the distribution function from its equilibrium
configuration. We
assume that the deviation is low enough, which we express
mathematically as the first order deviation from the equilibrium
distribution, denoted by
\begin{equation}
f(x,p,t)=f_0(p)+\delta f(x,p,t)
\end{equation}
where $f_0(p)$ is the equilibrium distribution, whereas $\delta
f(x,p,t)$ is small deviation from the equilibrium distribution of
mass points. The equilibrium distribution is assumed to be of
Maxwellian type, and for the sake of simplicity we assume
non-relativistic scenario with the Hamiltonian given by eqn(11a).
The Maxellian distribution of energy of the mass points is given by
\begin{equation}
f_0(p)=N_0 \exp\left [-\frac{\beta(x)\left (m_0c^2+\frac{p^2}{2m_0} \right
)}{k_BT}\right ]
\end{equation}
where $N_0$ is a constant depends on the temperature of the system.
Considering that the small perturbation is because of monochromatic type
gravity waves, we can write
\begin{equation}
\delta f(x,p,t) \propto \exp[i(kx-\omega t)]
\end{equation}
On substituting in Vlasov equation (eqn.(14)) the distribution function $f$, 
which is a sum of $f_0$,
the equilibrium part and $\delta f$, the small perturbation part, which
are given by eqns.(16) and (17) respectively, we have
\begin{equation}
-i\omega \delta f+ikv \beta(x)\left (1+\frac{p^2}{m_0^2c^2} \right
)^{-1/2} \delta f=m_0\alpha \left (1+ \frac{p^2}{m_0^2c^2}\right
)^{1/2} \frac{\partial f_0}{\partial p}
\end{equation}
Hence
\begin{equation}
\delta f=\frac{m_0\alpha \left ( 1+ \frac{p^2}{m_0^2c^2}\right
)^{1/2} \frac{\partial f_0}{\partial p}}
{i\left [ kv\beta(x) \left( 1+ \frac{p^2}{m_0^2c^2}\right )^{-1/2}
-\omega \right ]}
\end{equation}
where
\begin{equation}
\beta(x)=1+\frac{\alpha x}{c^2}
\end{equation}
Since the Rindler space is locally flat, we may use the results of
standard classical electrodynamics and can write the polarization
equation for gravity in the following form 
\begin{equation}
\frac{dP}{dx}=\delta \rho
\end{equation}
where $\delta \rho$ is the small change in matter density caused by the
incident gravity waves or in other wards a small deviation of mass 
distribution, given by
\begin{equation}
\delta \rho=m_0\int \delta f d^3p
\end{equation}
Assuming that the polarization function of the mass deviation function $P$ 
also varies harmonically and considering the
relation from the classical electrodynamics
\begin{equation}
P=\frac{(\epsilon_l-1)\alpha}{4\pi}
\end{equation}
we have
\begin{equation}
\epsilon_l =1+\frac{4\pi m_0^2}{k} \int  \frac{\left (1+
\frac{p^2}{m_0^2c^2}\right )^{1/2}} { \left [ \frac{kv\beta(x)}{\left (
1+ \frac{p^2}{m_0^2c^2}\right )^{1/2}} -\omega \right ] }
\frac{\partial f_0}{\partial p} d^3P
\end{equation}
where $\epsilon_l$ may be assumed to be the gravitational permittivity
or dielectric constant.
Defining
\begin{equation}
z^2=\frac{\beta(x)p^2}{2mk_BT} ~~{\rm{and}}~~
\gamma=\frac{k_BT}{\beta(x) c^2}
\end{equation}
we have
\begin{equation}
\epsilon_l=1+\frac{8\pi m_0^2 N}{k^2(k_BT\beta(x))^{1/2}}
\exp(-m_0/\gamma) \int_0^\infty \frac{(\gamma^2z^5+2\gamma z^3
+z)\exp(-z^2)}{\gamma uz^2-z+u} 
\end{equation}
where
\begin{equation}
u=\frac{\omega}{(k(2k_BNT\beta(x))^{1/2}}
\end{equation}
This is the general expression for the gravitational dielectric
constant as a function of frequency and wave number of the incident
gravity wave.
Now instead of going into all possible harmonics of the
incident gravity waves, for the sake of mathematical convenience,
we consider two extreme regions. 
The high frequency range, when $u\gg 1$, we have
\begin{equation}
\epsilon_l=1+A\int_0^\infty \frac{(\gamma^2z^5+2\gamma z^3 +z)\exp(-z^2)}{
\gamma uz^2 -z+u)} dz
\end{equation}
where
\begin{equation}
A=-\frac{8\pi Nm_0^2}{k^2(2k_B\beta(x))^{1/2}} \exp(-m_0/\gamma))
\end{equation}
On integrating over $z$, we get
\begin{equation}
\epsilon_l= 1+2\pi iA \left [ \frac{(1-3\gamma u^2\gamma^2
u^4)}{\gamma^2u^4} +\frac{2(1-\gamma u^2)}{\gamma u^2}+1 \right ]
\exp \left ( -\frac{1-2\gamma u^2+(1-4\gamma u^2)^{1/2}} {2\gamma^2
u^2}\right )
\end{equation}
Now in the high frequency range, it is quite possible that the
quantity within the square root in the exponential term may become
negative. Then the permittivity may be written as
\begin{equation}
\epsilon_l=\epsilon_l^{(R)}+\epsilon_l^{(I)}
\end{equation}
where $\epsilon_l^{(R)}$ and $\epsilon_l^{(I)}$ are the real and
imaginary components of the permittivity $\epsilon_l$ in the high
frequency range and are given by
\begin{equation}
\epsilon_l^{(R)}= 1+2\pi A \left [ \frac{(1-3\gamma u^2\gamma^2
u^4)}{\gamma^2u^4} +\frac{2(1-\gamma u^2)}{\gamma u^2}+1 \right ]
\exp \left ( -\frac{1-2\gamma u^2} {2\gamma^2
u^2}\right )\sin\left [\frac{(4\gamma u^2-1)^{1/2}}{2\gamma^2 u^2}
\right ]
\end{equation}
and 
\begin{equation}
\epsilon_l^{(I)}= 1+2\pi A \left [ \frac{(1-3\gamma u^2\gamma^2
u^4)}{\gamma^2u^4} +\frac{2(1-\gamma u^2)}{\gamma u^2}+1 \right ]
\exp \left ( -\frac{1-2\gamma u^2} {2\gamma^2
u^2}\right )\cos\left [\frac{(4\gamma u^2-1)^{1/2}}{2\gamma^2 u^2}
\right ]
\end{equation}
respectively

In fig.(1) we have shown the variation of the ratio of real part of 
$\epsilon_l$ and $A$ with
the frequency of the radiation. It has been observed from this figure 
that the real
part becomes exactly one for the the large values of frequency.
Since $A$ is independent of frequency, we may conclude that the
gravitational dielectric constant of the medium for gravity waves in
this region is $\propto A$.
\begin{figure}[ht]
\psfig{figure=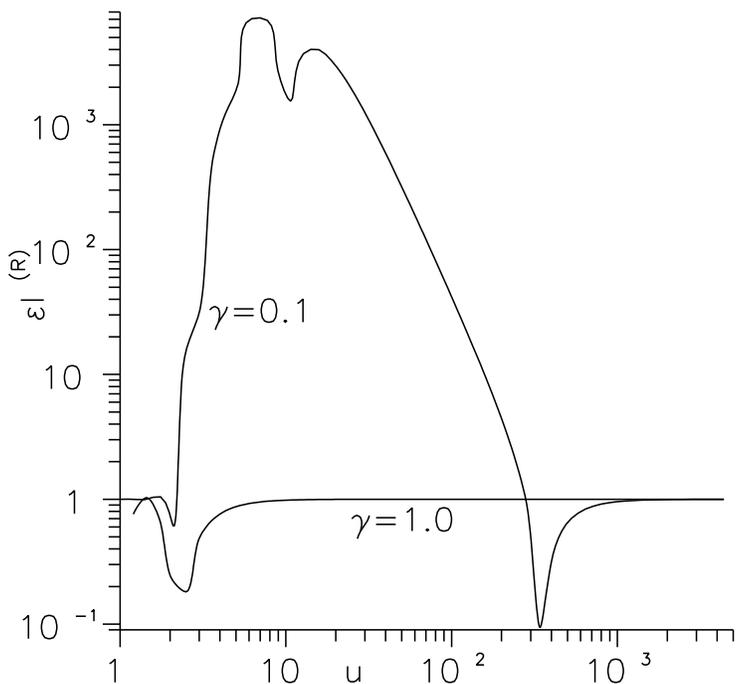,height=0.5\linewidth,angle=0}
\caption{The variation of the scaled  real part of $\epsilon_l$
with frequency in the high frequence region}
\end{figure}
In fig.(2) we have shown the frequency dependence of the corresponding
imaginary part of permittivity. It is evident from this figure that the
imaginary part becomes extremely small for the high
value of radiation frequency.
\begin{figure}[ht]
\psfig{figure=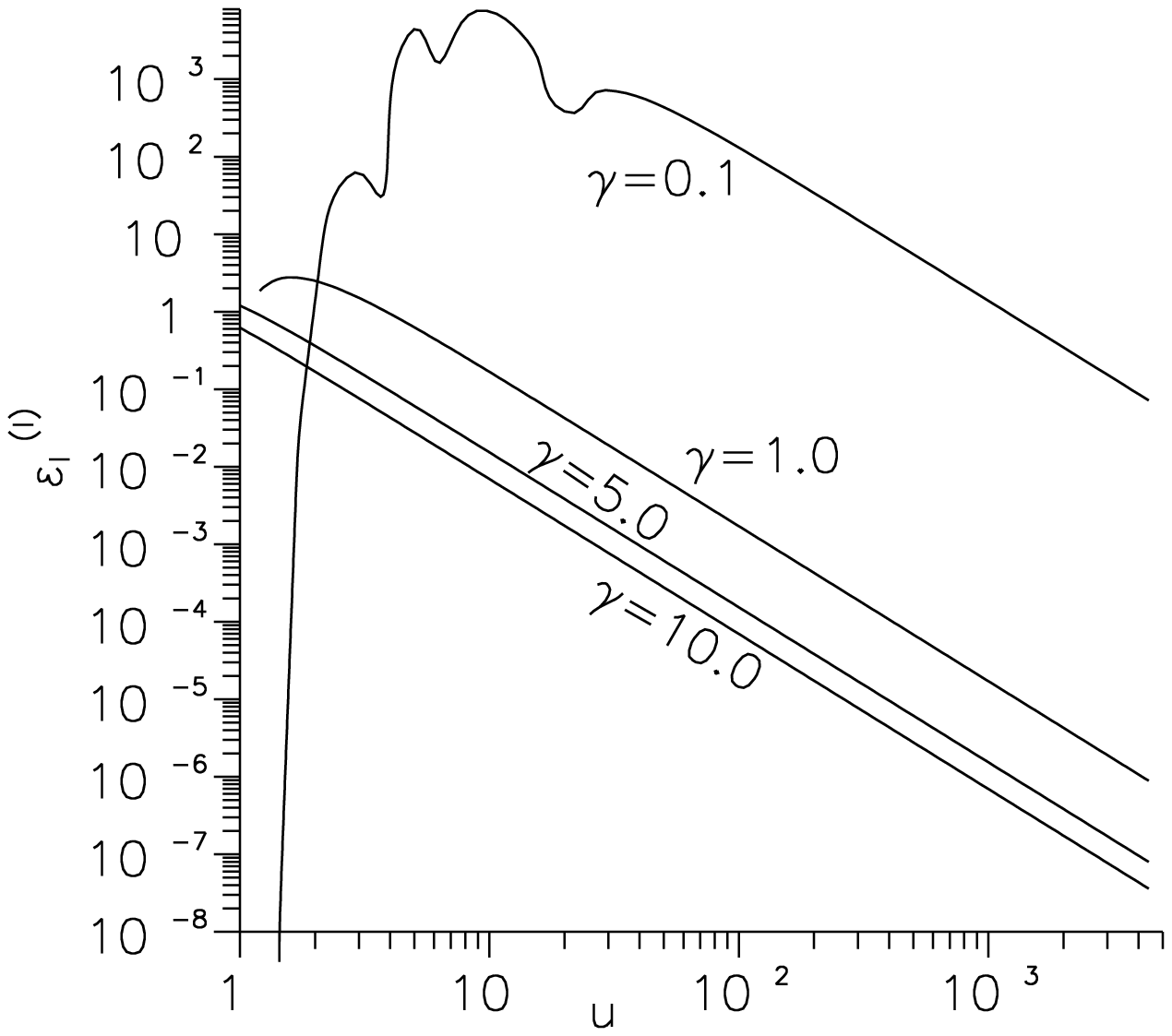,height=0.5\linewidth,angle=0}
\caption{The variation of the scaled  imaginary part of $\epsilon_l$
with frequency in the high frequence region}
\end{figure}
Further, the actual value of the gravitational permittivity becomes 
exactly one, a real quantity for high values of
$\alpha$. This is because as gravitational field 
$\alpha \longrightarrow \infty$, the quantity
$\gamma\longrightarrow 0$, then the exponential term present in the
expression for permittivity makes it
exactly equal to one. Therefore we expect that the
region very close to the event horizon of a black hole, where the
gravitational field is extremely high, is
transparent for the propagation of gravity waves. Therefore, there will
be no gravitational bending of plane gravity waves near strongly
gravitating objects. On the other
hand, it is well known that there are gravitational bending of
electromagnetic waves while pass through strong gravitational field.
Further, it has been shown that 
free space behaves like a refracting medium in presence
of gravity \cite{R1,SOSO}. The Rindler space, which is locally flat
also behaves like a refracting
medium. Further, it has been observed that since strong
gravitational field makes the refracting index quite high, the speed
of light decreases with the increase in gravitational field. Of
course the gravitational refractive index for optical waves remains
always real. On the other hand for gravity waves there is no such effect
of strong gravitational field. The physical reason behind such
non-interacting nature of gravity waves and strong gravitational field
is as follows: It is well known that the propagation of gravity waves is
the fluctuation of space-time coordinates at different points from 
their flat configuration. It is also well established that the
distortion of space-time geometry near a strongly gravitating object is
quite high and since in this case during the propagation of 
gravity wave, where the space time fluctuation  effect is
negligibly small compared to such strong
distortion of space-time geometry, we may
conclude that there is almost no interaction of gravity waves with
the gravitational field of strongly gravitating objects.

We next consider the low frequency gravity waves, i.e., $u\ll 1$.
In this case the permittivity becomes
\begin{equation}
\epsilon_l=\epsilon_l^{(R)}+i\epsilon_l^{(I)}
\end{equation}
where the real part is given by
\begin{equation}
\epsilon_l^{(R)}=1
\end{equation}
and is independent of frequency. Therefore it is always unity.
The corresponding imaginary part is given by
\begin{eqnarray}
\epsilon_l^{(I)}&=&2\pi A[-2\gamma^2u^6+\gamma^2 u^5
+(5\gamma^2-14\gamma)u^4 +2\gamma u^3 
+(16\gamma -34)u^2\nonumber \\ &+&u  
+ \left (\frac{4}{\gamma}-\frac{14}{\gamma^2}
\right )
\frac{1}{u^2} + \left ( \frac{1}{\gamma^2}- 
\frac{2}{\gamma^3}
\right )\frac{1}{u^4} -\frac{34}{\gamma} +17] \exp(-1/\gamma^2
u^2)
\end{eqnarray}
Since the real part of the permittivity in low frequency region is
always one, we have plotted the variation of imaginary part with
frequency in fig.(3) for various values of $\gamma$. Here also we have
actually plotted the ratio of the imaginary part of permittivity and
$A$. In this case
the imaginary part saturates to the value $\sim 100$ for relatively
high frequency range. Since the imaginary part is quite large,
we expect that there will be reasonable amount of absorption of
low frequency radiation in the medium, i.e., in the plasma of point
galaxies and the low frequency part of gravity waves may not be possible
to observe.
\begin{figure}[ht]
\psfig{figure=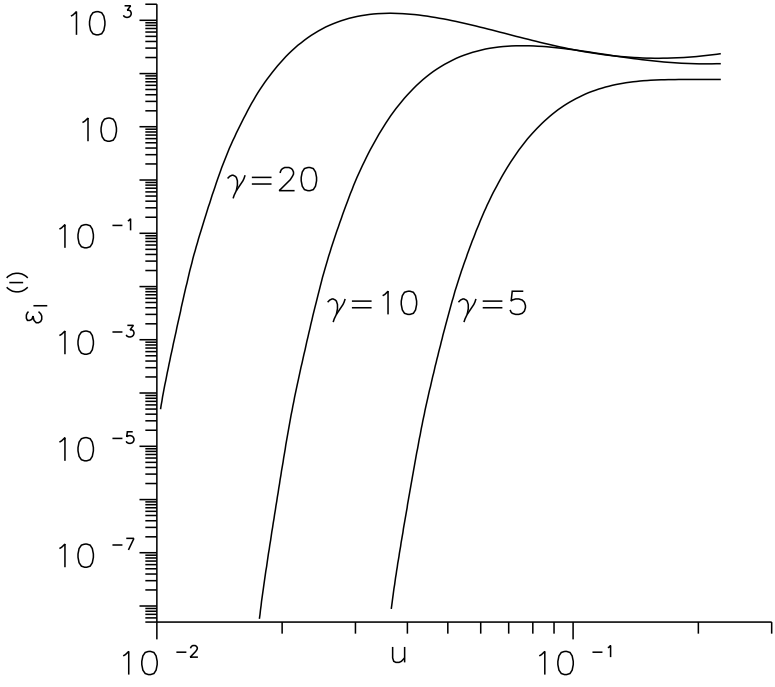,height=0.5\linewidth,angle=0}
\caption{The variation of the scaled  imaginary  part of $\epsilon_l$
with frequency in the low frequency region}
\end{figure}
\section{Boltzmann Equation for Stellar Plasma in Rindler Space}
We next consider a collisional stellar plasma in Rindler space. We
put the collision term by hand on the right hand side of the Vlasov
equation and is given by
\begin{equation}
\frac{\partial f}{\partial t}-v\beta(x)\left (1+
\frac{p^2}{m_0^2c^2}\right )^{-1/2} \frac{\partial f}{\partial x}
-m_0\alpha \left ( 1+\frac{p^2}{m_0^2c^2}\right )^{1/2}
\frac{\partial f}{\partial p}=C[f]
\end{equation}
This is the Boltzmann equation in Rindler space. 
Since it is not possible to evaluate the collision term from the
cross section of the elementary processes, we make relaxation time
approximation, given by
\begin{equation}
C[f]=-\frac{f(x,p,t)-f_0(p)}{\tau}
\end{equation}
where $\tau$ is the relaxation time and it is assumed that the
system is very close to the equilibrium configuration. The  
function $f_0(p)$ is the equilibrium distribution function. For the
evaluation of kinetic coefficients, for the sake of simplicity we
assume that $f_0(p)$ is Maxwellian in nature and in 
the present scenario the equilibrium distribution is given by
\begin{equation}
f_0(p)=C\exp\left (\frac{\mu-\varepsilon(x,p)+Vp}{k_BT}\right )
\end{equation}
where $\mu\equiv \mu(x,t)$, the local chemical potential, $T\equiv
T(x,t)$, the local temperature and $V\equiv V(x,t)$, the local flow
velocity, which are also changing with time and the single particle
energy in the non-relativistic approximation is given by
\begin{equation}
\varepsilon(x,p)=\beta(x)\left (\frac{p^2}{2m_0}+m_0c^2\right )
\end{equation}
Here our intention is to obtain the kinetic coefficients of the stellar
plasma in Rindler space. In doing so we proceed as follows:
Now keeping only the
equilibrium distribution function  on the left hand side of Boltzmann
equation, we have
\begin{equation}
\left [ \frac{\partial}{\partial t}+ v\beta(x)
\frac{\partial}{\partial x}- m_0\alpha (1+\frac{p^2}{2m_0^2c^2}
\frac{\partial}{\partial p}\right ]f_0(p)=C[f]
\end{equation}
where $C[f]=-\delta f(x,p,t)/\tau$ in linear approximation, where $\delta
f(x,p,t)$ is the small deviation of distribution function from the
equilibrium configuration. 
Then following Huang \cite{HU} for the usual version of
Boltzmann equation and taking into account
\begin{equation}
\frac{\partial f_0}{\partial \varepsilon}=-\frac{f_0}{k_BT}
\end{equation}
we have 
\begin{eqnarray}
&&\big [ \frac{(\varepsilon -h)\beta(x)}{T}v\frac{\partial
T}{\partial x}+ m_0\beta(x)u_xu_y\zeta_{x,y}+ m_0\beta(x)\left (
1+\frac{p^2}{2m_0^2c^2}\right )\alpha
v\nonumber \\ &+&m_0(1-\beta(x))v\frac{\partial v}{\partial t}
-\frac{\varepsilon
-h+TC_p}{C_v/k_B}\frac{\partial u_x}{\partial x}\big
]\frac{f_0}{k_BT}=-\frac{\delta f}{\tau}
\end{eqnarray}
In the left hand side there are all possible driving forces, e.g.,
thermal conduction, viscous flow (both shear and bulk) and flow
under gravity. Here
\begin{equation}
\zeta_{xy}=\frac{1}{2}\left( \frac{\partial u_x}{\partial z}
+\frac{\partial u_z}{\partial x} \right )
\end{equation}
is the shear tensor. The term with the driving force $\partial
u_x/\partial x$ represents the bulk viscosity. In our present study
we discard this term. Our next task is to obtain the three
different types of transport coefficients, viz, thermal
conductivity, shear viscosity coefficient and coefficient of
gravitational flow.
\section{Shear Viscosity Coefficient}
To obtain the shear viscosity coefficient we consider only the
driving force associated with shear flow. The relevant form of the
deviation of distribution function from its equilibrium structure
is given by
\begin{equation}
\delta f=-\frac{\tau}{k_BT}\frac{m_0\beta(x)} {\left ( 1+
\frac{p^2}{m_0^2c^2}\right )}   u_xu_y\zeta_{xy}
\end{equation}
Now from the definition of viscus flow we can write down the
relevant form for the expression of pressure tensor in the following form
\begin{equation}
\Pi_{xy}=n\int d^3p p_xv_z (f_0(p)+\delta f(x,p,t))
\end{equation}
where $n$ is the number density and the above equation basically
indicates the momentum transfer along $z$-direction, while the
actual flow is along $x$-direction. Now it a matter of simple
algebra to show that in the expression for pressure tensor, the
equilibrium part does not contribute. This is also evident from the
physical nature of equilibrium configuration. Therefore after
integrating over momentum in the rest frame of the fluid element when
$u\rightarrow v$ for all the components
and comparing with equation for viscous flow, given by
\begin{equation}
\Pi_{xy}=-\eta \frac{du_x}{dz}
\end{equation}
we have for the coefficient of shear viscosity
\begin{equation}
\eta=\frac{n\tau k_BT}{\beta^4(x)} \exp\left (
-\frac{2m_0c^2\beta(x)}{k_BT} \right )
\end{equation}
It is obvious that for $\alpha=0$, i.e., in absence of
gravitational field when $\beta=1$, we get back the conventional
result in absence of external gravitational field, or in flat space-time. 
Whereas for
ultra strong gravitational field, i.e., for $\alpha \longrightarrow
\infty$ or $\beta \longrightarrow \infty$ the
coefficient of viscosity becomes vanishingly small.
\section{Thermal Conductivity}
In this case the thermal current is given by
\begin{equation}
j_\epsilon^x=\int d^3p \varepsilon v_xf(x,p,t)
\end{equation}
Again the equilibrium part does not contribute. Therefore we have
\begin{equation}
j_\epsilon^x=\int d^3p \varepsilon v_x \delta f(x,p,t)
\end{equation}
Further the relevant form of the deviation part of distribution
function is given by
\begin{equation}
\delta f(x,p,t)=\frac{\tau f_0(p)}{k_BT}\frac{(\epsilon -h)
\beta(x)}{T} u \frac{dT}{dx}
\end{equation}
Substituting this value in the thermal current and considering the
standard definition of thermal current
\begin{equation}
j_\epsilon^x=-K\frac{dT}{dx}
\end{equation}
we have after integrating over momentum in the local rest frame
\begin{eqnarray}
K&=&\frac{n\tau}{4m_0 T\beta^{3/2}(x)} \exp\left (
-\frac{m_0c^2\beta(x)}{k_BT}
\right )\\ \nonumber &&  \left [ \frac{35k_B^2T^2}{\beta(x)}+ 10k_BT
m_0c^2 +\left (\frac{10k_BT}{\beta(x)} +m_0c^2\right ) (m_0c^2
\beta(x)- \frac{5}{2} k_BT)\right ]
\end{eqnarray}
Exactly like the shear viscosity coefficient, the thermal
conductivity also reduces to its standard form for $\beta=1$ and
vanishes for large gravitational field.
\section{Coefficient of Gravitational Flow}
This is the first time such coefficient is obtained. We are the
first to calculate this quantity. The relevant form of Boltzmann
equation in this case is given by
\begin{equation}
\left [ n_0(1-\beta(x))v\frac{dv}{dt} +m_0\beta(x) \left (
1+\frac{p^2}{2m_0^2c^2}\right )\alpha v\right ] \frac{f_0(p)}{k_BT}
=-\frac{\delta f}{\tau}
\end{equation}
From the equation of motion we have
\begin{equation}
m_0(1-\beta(x))v\frac{dv}{dt}=-m_0(1-\beta(x))\left (
1+\frac{p^2}{2m_0 c^2} \right )\alpha v
\end{equation}
Hence the deviation part of the distribution function may be
written as
\begin{equation}
\delta f=\frac{(1-2\beta(x))m_0\tau}{k_BT}\left (1+
\frac{p^2}{2m_0^2c*2} \right )\alpha v
\end{equation}
Now we may define the mass flow current under the influence of gravity
in the following form
\begin{equation}
j_m(x)=m_0\int d^3p v_x f(x,p,t)
\end{equation}
Assuming again $f(x,p,t)=f_0(p)+\delta f(x,p,t)$ and since $f_0(p)$
does not contribute, we have
\begin{equation}
j_m(x)=m_0\int d^3p v_x \delta f(x,p,t)
\end{equation}
On substituting $\delta f(x,p,t)$ in the above expression,
integrating over momentum in the local rest frame and 
finally from the analogy of charge current flow, we can write
\begin{equation}
j_m(x)=\sigma \alpha ~(j=\sigma E ~ {\rm{in ~the ~ electrical~ case}})
\end{equation}
we have the coefficient for the gravitational flow of mass
\begin{equation}
\sigma=\frac{n\tau (1-2\beta(x))}{3c^2 \beta^{5/2}} \left ( \frac{
5k_BT}{2} +m_0c^2\right ) \exp \left (-\frac{m_0c^2\beta(x)}
{k_BT}\right )
\end{equation}
Since $(1-2\beta(x))=-(1+\frac{2\alpha x}{c^2})$, a negative
quantity, the coefficient of mass flow under gravity is also
negative. The physical reason for negative value is because unlike
all other kind of flows, where current travels from high potential to low 
potential region, the mass flow occurs from low gravity region
to high gravity region. Further like all other kinetic coefficients
this particular kinetic coefficient also vanishes for high
gravitational field. However, the formalism to obtain this kinetic
coefficient is not valid for  zero gravitational field. Therefore the
stellar matter in presence of ultra strong gravitational field behaves
like an ideal fluid. Which may occur at the vicinity of event horizon.
This may also be true if the matter is created near the event horizon of
a black hole through Hawking radiation \cite{HW}. According to principle
of equivalence the uniform acceleration of the frame equivalent to a
constant gravitational field of a rest frame, then the radiation or
material particles observed in quantum vacuum by the uniformly
accelerated frame as Unruh effect \cite{UN} 
will also behave like an ideal fluid.
\section{Conclusion}
We have seen that in the high frequency region of gravity wave, the
real part of gravitational permittivity 
converges to unity at relatively higher frequency values,
whereas the imaginary part becomes extremely small. We have also
observed that the actual value of the permittivity becomes exactly
one, a real number for very high gravitational field, which makes
the medium transparent for gravity waves. On the other hand in the
low frequency region, the real part is always unity, whereas the
imaginary part saturates to a value $\sim 100$, which makes the
medium strongly absorbing in that frequency zone.


\begin{thebibliography}{99}
\bibitem{R1} Landau L.D. and Lifshitz E.M., The Classical Theory 
of Fields, 
Butterworth-Heimenann, Oxford, (1975).
\bibitem{R2} W.G. Rosser, Contemporary Physics, {\bf{1}}, 453, (1960).
\bibitem{R3} N.D. Birrell and P.C.W. Davies,
Quantum Field Theory in Curved Space, Cambridge University Press,
Cambridge, (1982).
\bibitem{R31} Torres  del Castillo G.F. and Perez 
Sanchez C.L.,  Revista Mexican De Fisika 52, 70, (2006).
\bibitem{R32} M. Socolovsky, Annales de la
Foundation Louis de Broglie {\bf{39}}, 1, (2014).
\bibitem{R4} C.G. Huang and J.R. Sun, arXiv:gr-qc/0701078, (2007).
\bibitem{R5} Domingo J Louis-Martinez, Class. Quantum Grav.,
{\bf{28}}, 036004, (2011). 
\bibitem{R6} D. Percoco and V.M. Villaba, Class. Quantum Grav.,
{\bf{9}}, 307, (1992).
\bibitem{R7} S. De, S. Ghosh and S. Chakrabarty, Astrophys and Space
Sci, 360:8, DOI 10.1007/s10509-015-2520-3. (2015).
\bibitem{R71} S. De, S. Ghosh and S. Chakrabarty, Mod. Phys. Lett. A
{\bf{30}}, 1550182 (2015).
\bibitem{SOSO} Soma Mitra and Somenath Chakrabarty (to be submitted,
2017).
\bibitem{LP} Physical Kinetics, E,M. Lifshitz and L.P. Pitaevskii, 
Vol.10, Butterworth-Heimenann, Oxford, (1998).
\bibitem{HU} Statistical Mechanics, K. Huang, Wiley, NY, 1965.
\bibitem{HW} S.W. Hawking,  Nature, {\bf{248}}, 30, (1974);
S.W. Hawking,  Comm. Math. Phys.  {\bf{43}}, 199, (1975).
\bibitem{UN} W.G. Unruh,  Phys. Rev. {\bf{D14}}, 4, (1976);
W.G. Unruh, Phys. Rev. {\bf{D14}}, 870, (1976).
\end{thebibliography}
\end{document}